\documentstyle[twocolumn,aps,psfig]{revtex}

\begin{document}

\bibliographystyle{unsrt}

\title{Studies of $h/e$ Aharonov-Bohm Photovoltaic Oscillations in 
Mesoscopic Au Rings}
\author{R. E. Bartolo$^*$ and N. Giordano} 
\address{Department of Physics, Purdue University, 
West Lafayette, IN 47907-1396}
\vspace{.25in}
\author{X. Huang$^\dagger$ and G. H. Bernstein} 
\address{Department of Electrical Engineering, University of Notre Dame, 
Notre Dame, IN 46556}
\maketitle
\begin{abstract}
We have investigated a mesoscopic photovoltaic (PV) effect in
micron-size Au rings in which a dc voltage $V_{dc}$ is generated 
in response to microwave radiation. The effect is due to the 
lack of inversion symmetry in a disordered system.
Aharonov-Bohm PV oscillations with flux period $h/e$ have been 
observed at low microwave intensities for temperatures ranging 
from 1.4 to 13 K.  For moderate microwave intensities the 
$h/e$ PV oscillations are completely quenched providing 
evidence that the microwaves act to randomize the phase 
of the electrons. Studies of the temperature dependence of $V_{dc}$ 
also provide evidence of the dephasing nature of the microwave field. 
A complete theoretical explanation of the observed behavior seems to 
require a theory for the PV effect in a ring geometry.   
\end{abstract}

\section{Introduction}

Studies of the transport properties of disordered mesoscopic
systems have revealed a number of interesting quantum
interference phenomena \cite{lee90}. These effects include the
observation
of Aharonov-Bohm  or $h/e$ conductance oscillations in micron-size Au
rings, universal (or aperiodic) conductance fluctuations (UCF) in small
wires, and the time-dependent conductance fluctuations due to the motion
of even  a single defect or impurity. In each case, the conductance
exhibits  fluctuations  of order $e^2/h$,  provided the electron retains
phase coherence over the length of the system.
Central to the observation of such interference phenomena is the fact
that an electron scatters coherently off the random impurity configuration
over a distance, $L_\phi$, defined as the phase coherence length.
Another consequence of coherent impurity scattering is that a
disordered mesoscopic  system lacks a center of inversion symmetry which
allows for non-linear effects that would otherwise be forbidden.

One such non-linear response concerns a mesoscopic photovoltaic (PV)
effect \cite{falko89a,falko89b} in which a dc-voltage, $V_{dc}$, is
generated in response to microwave radiation, even with no current
applied from external leads. The PV effect has much in common with UCF
and consequently $V_{dc}$ is sensitive to  small changes in the
impurity distribution and magnetic field, $H$. Initial PV experiments
centered on studies of small microbridges  where $V_{dc}$ was shown to
exhibit aperiodic fluctuations as a function of $H$ at low temperatures
($\sim$ 4 K) \cite{liu92,bykov90}.  More recent studies of the
PV effect in submicron-size rings  clearly revealed Aharonov-Bohm PV
oscillations  of flux period $h/e$, in addition to the aperiodic
fluctuations \cite{bykov93,bartthesis}. One surprising feature
of these results was that the PV signal was quite  pronounced,
even for temperatures at which the analogous conductance fluctuations
were minute and difficult to observe.
The robust nature of these results \cite{liu92,bykov90,bykov93,bartthesis} 
demonstrates that the mesoscopic PV effect also promises to be a powerful 
and convenient probe of mesoscopic systems and may lead the way for
possible microwave device applications.

In this paper we report detailed studies of the PV effect as a function
of magnetic field, microwave power, and temperature in
submicron-diameter
Au rings.  Studies of $V_{dc}$ as a function of the microwave power show
that the $h/e$ PV oscillations are quenched in response to a moderate
amount of microwave radiation. Our data suggests that the  microwave
period plays a fundamental role in determining the temperature
dependence of the PV response when it is of order the phase coherence
time, $\tau_\phi$. Although a complete analysis of our results awaits an
extension of the PV theory  to include the case of a ring geometry, our data
provides evidence of the dephasing nature of a high-frequency
field in a mesoscopic system. 

\section{Theory of the Mesoscopic Photovoltaic Effect}

\subsection{The Mesoscopic Photovoltaic Effect At Zero Temperature}

The calculation of Fal'ko and Khmel'nitskii
\cite{falko89a} for the magnitude of the PV effect considers the case of
a metallic mesoscopic wire or microbridge illuminated by high-frequency
(microwave) radiation. A conductor in the presence of a high frequency
field $E_{ac}$ will absorb
\begin{equation} N_e={E_{ac}^2 L^2 G \over \hbar \omega}
\label{eq:N_e}
\end{equation}
photons per unit time, where $G$ is the conductance,
$L$ is the sample length (where $L \leq L_\phi$), and $\hbar \omega$ is
the photon energy. In the weak-field limit (i.e., $ eE_{ac}L/\hbar \omega
\ll 1$), an electron will absorb at most one photon, and $N_e$ equals
the number of electrons that are excited into empty states above the
Fermi energy, $E_F$. If the scattering potential of the electron is perfectly
symmetric then  half the carriers will diffuse to the left voltage probe 
and half to the right. In this case the net photovoltage would be zero. 
However, due to the asymmetry of the random impurity potential, a different 
number of electrons diffuse to opposing voltage contacts.  The degree of 
asymmetry $\alpha$ in a disordered mesoscopic junction was estimated  by the
theory \cite{falko89a} as the ratio of the magnitude of the UCF to the sample
conductance: $ \alpha \sim e^2/hG $.  Putting these effects together,
the net dc voltage $V_{dc}$ generated by the microwaves is given by
$V_{dc} \sim (eN_e\alpha)R$, where $R$ is the sample resistance. In the
low-frequency limit ($\omega \tau_f \ll 1$) the relevant time scale of
the system is determined by $\tau_f$, the time it takes the electron to
diffuse to the voltage contacts, not the period ($\sim \omega^{-1}$) of
the microwave field. In this case,
$\tau_f$ is  substituted for $\omega^{-1}$ in Eq.~\ref{eq:N_e} and the
magnitude of the net microwave induced dc voltage is given by
\begin{equation} \overline{V_{dc}} \approx  {e \over \tau_f} R
 \left( {eE_{ac}L\tau_f \over \pi^2 \hbar}\right)^2
\left[{\lambda V_d \over L^d }\right]^{1/2}, ~\omega\tau_f \ll 1.
\label{eq:Vdc1} \end{equation}
The numerical factors $\lambda$, and $V_d$, the volume of the junction,
are given from the quantitative calculation and depend on the sample
dimensionality $d$ (see Table 1 in Ref. 2). Note that
$\overline{V_{dc}}$ in Eq.~\ref{eq:Vdc1} is the rms voltage
averaged over different realizations of the random potential.

The deep connection between the PV effect and UCF hinges on the fact
that the degree of asymmetry is related to the electron's transmission 
probability across the sample. As a result, the magnitude and polarity of 
$V_{dc}$ is a random function 
of the magnetic field $H$, the energy of the electron, and the precise 
location of impurities. The field change required to effect a full 
fluctuation in $V_{dc}$ is defined as the correlation field, 
$H_c \sim \Phi_o /L_\phi w$, 
and is of order the field required to produce one quantum of flux
($\Phi_o$) through a phase-coherent area $L_\phi \times w$,
where $w$ is the width of the wire \cite{stone87}. 
Varying the field strength by more than
$H_c$ causes a shift in the electron phases that is equivalent to a 
complete rearrangement of the impurity configuration, thus causing 
aperiodic PV fluctuations, $\overline{V_{AF}}$, that are of order 
$\overline{V_{dc}}$ in Eq.~\ref{eq:Vdc1}.  

The energy scale over which the electrons are correlated is given, 
in analogy with UCF, by the Thouless energy 
$E_c \sim \hbar / \tau_f$ \cite{lee90,stone87}. 
When the spread of the electron's energy distribution ($\Delta E$) is 
greater than $E_c$,
the net current is averaged statistically due to the superposition
of $n=\Delta E /E_c$ uncorrelated energy intervals. Therefore, 
Eq.~\ref{eq:Vdc1} represents the most optimistic estimate for the PV 
effect as it does not include the
adverse averaging effects due to finite temperatures (i.e., it was 
assumed that $L_\phi \sim L$ and $k_BT<E_c$). Microwave
frequencies are optimal for studies of the PV effect since the associated 
photons are energetic enough to excite a large number of electrons into 
empty states above the Fermi energy, while not leading to excessive energy
averaging (i.e. $\Delta E = \hbar \omega \sim E_c$). For a typical
metallic mesoscopic sample, the degree of asymmetry is small
($\alpha \sim 10^{-3}$) and $\overline{V_{dc}}$ is on the order of
1 nV at a frequency of 10 GHz for $E_{ac}=$1 V/m and
$R=$10 $\Omega$. Note that the PV effect is entirely
mesoscopic in  origin and there is  no inherent macroscopic or
background component  as in the case of the UCF. In a disordered
macroscopic system (i.e. $L_\phi \ll L$)
the PV effect is negligibly small since currents from individual
phase-coherent regions are random in sign and add incoherently. 

\subsection{The Mesoscopic Photovoltaic Effect in the High-Frequency
and High-Power Limits}

The microwave field can act to randomize the phase of the
electron at sufficiently high powers.
Previous theoretical \cite{altshuler81} and experimental
\cite{lindelof87,kvon88,liuthesis,kuchar91} studies have centered on the
dephasing effects of microwaves on the weak localization 
magnetoresistance. The experiments are complicated by electron heating
effects, which can also be important at high power levels
\cite{kuchar91}. Nevertheless, phase breaking due to the ``dynamical''
influence of microwaves (apart from heating) has been observed
\cite{lindelof87,kvon88,liuthesis}. The dephasing effects of a microwave
field on the $h/e$ oscillations in a mesoscopic system have yet to be
studied. In related work \cite{price94}, the effect of a finite
measurement frequency (from 250 to 1.2 GHz) on the complex magnetoconductance 
of submicron-diameter Ag rings has been studied, although no suppression of
the $h/e$ conductance amplitude was observed.
As we will discuss below, it is possible to study the dephasing effects
of a microwave field through a detailed analysis of the microwave power
and temperature dependence of the $h/e$ PV oscillations in samples with
a ring geometry.

The microwave power and temperature dependence of $\overline{V_{dc}}$
depends critically on the interplay between the relative magnitudes of
the strength ($E_{ac}$) and frequency ($\omega$) of the microwaves, and the
mesoscopic scales of time ($\tau_\phi$ and $\tau_f$) and energy
($E_c$) discussed above.  The theory \cite{falko89a} calculates the 
magnitude of the PV effect in the various asymptotic regimes of the 
irradiating field parameters. 
In the low-power high-frequency limit (i.e., $\hbar \omega > E_c$ and
$ \omega^{-1} < \tau_f$), $V_{dc}$ is reduced in magnitude by a factor of 
$n^{-1/2}$ due to photon energy averaging 
(where  $n= \hbar \omega /E_c = \pi^{-2}\tau_f \omega
$). Eq.~\ref{eq:Vdc1} is also modified by the substitution
$\tau_f \rightarrow \omega^{-1}$ and $V_{dc}$ is given by 
\cite{falko89a}
\begin{equation}  \overline{V_{AF}} \approx eR \left({\omega \over
\tau_f}\right)^{1\over 2} \left({eE_{ac}L \over \hbar \omega}\right)^2
\left[\lambda{V_d \over L^d}\right]^{1/2}, ~\omega \tau_f \gg 1.
\label{eq:Vdc4} \end{equation}
Note that in the high-frequency limit $V_{dc}$ is strongly frequency
dependent, in comparison to Eq.~\ref{eq:Vdc1} which is independent of
frequency.

In the limit of strong microwave fields the
electron may absorb more than one photon in time $\tau_f$ which
leads to energy and phase relaxation of the carriers \cite{falko89a}.
The number of photons $\xi^2$ absorbed by a particular electron during
a diffusion time $\tau_f$ is given by
\begin{equation} \xi^2 = {N_e \tau_f \over N}=
\left({eE_{ac}L\over \hbar \omega}\right)^2,
\label{eq:photonabs} \end{equation}
where $N_e$ is defined by Eq.~\ref{eq:N_e}, $N=(dN/dE)\hbar \omega$
is the total  number of  electrons within $\hbar \omega$ of the Fermi
energy, and the conductance $\sigma $ is defined by 
$\sigma V_d=e^2D(dN/dE)=GL^2$ (where $D$ is the diffusion  constant).
Assuming the electron absorbs at least one additional photon
before exiting the sample (i.e., $\xi^2=1$),  a microwave-enhanced
phase relaxation time $\tau_{ac}$  can easily be determined from
Eq.~\ref{eq:photonabs},
\begin{equation} \tau_{ac} = {N \over N_e} =  { (\hbar \omega)^2 \over
(eE_{ac})^2 D}.
\label{eq:tau_ac} \end{equation} The length scale over which the
microwaves randomize the electron's phase is defined by
$L_{ac}=(D\tau_{ac})^{1/2}=\hbar \omega/eE_{ac}$. 
Although the connection was not made explicit
by the PV theory \cite{falko89a}, $\tau_{ac}$ has been studied in
connection with the dephasing effects of a microwave
field on the weak localization magnetoresistance
\cite{altshuler81,lindelof87,kvon88,liuthesis,kuchar91}.

For strong microwave fields ($\xi \gg 1$) the theory
\cite{falko89a} predicts that the linear power dependence of
$\overline{V_{AF}}$ in Eqs.~\ref{eq:Vdc1} and \ref{eq:Vdc4} will
saturate due to energy and phase relaxation of the carriers.  As a
result, the quadratic field dependence is replaced by a much weaker 
logarithmic dependence \cite{falko89a}.
$V_{dc}$ in the high-power high-frequency limit is given by
\begin{equation} \overline{V_{AF}} \approx \sqrt{20}eR\left({\omega
\over \tau_f}\right)^{1/2} \ln ( \xi ), ~~\omega \tau_f \gg 1, 
\label{eq:region3pow} \end{equation}
where $L >> L_{ac},w,t$ for a 1D sample of width $w$ and thickness $t$.
The logarithmic power dependence results from a competition between
the dephasing effect of the microwaves and
the growth of the number of microwave generated carriers $N_{e}$ with
increasing power. Therefore, the effects of a microwave-enhanced
phase relaxation time ($\tau_{ac}$) will be apparent in measurements of
$V_{dc}$ over a wide range of microwave powers.

\subsection{The Effect of Partial Phase Coherence and Thermal Energy 
Averaging}

For high temperatures, $L_\phi$ can be smaller than $L$, and $V_{dc}$ 
decreases with increasing temperature due to the averaging of 
$n=L/L_\phi$ phase coherent subsystems.
Quantitative estimates for $\overline{V_{AF}}$ as a function of 
temperature follow naturally from
Eqs.~\ref{eq:Vdc1}, \ref{eq:Vdc4}, and \ref{eq:region3pow} using the 
substitution $L \rightarrow L_\phi$, $\tau_f \rightarrow \tau_\phi$, 
and the relation $\tau_\phi \rightarrow L_\phi^2/D$.
The temperature dependence in the high-power high-frequency regime
is given \cite{falko89a} from Eq.~\ref{eq:region3pow}
\begin{equation}
\overline{V_{AF}}(T)\approx
\sqrt{20} eR\left({\omega \over \tau_f}\right)^{1/2}\ln[\xi(T)],
\label{eq:tempdep2} \end{equation}
where the parameter $\xi(T) = eE_{ac}L_\phi/ \hbar \omega $
varies with temperature \cite{falkonote2} since 
$L_\phi \sim T^{-p/2}$ (where $2/3<p<3$ depending on the phase relaxation
mechanism and the sample dimensionality \cite{altshuler81}). 
The precise temperature dependence depends on the relative value of
$\tau_\phi$ and $\omega^{-1}$. For temperatures greater than
around 5 K, $\tau_\phi$ can become less than $\omega^{-1}$, and the
temperature dependence becomes much stronger due to the
substitution $\omega^{-1} \rightarrow \tau_\phi$. We would
therefore expect $\overline{V_{AF}}$ to vary logarithmically with temperature.
The parameter $\xi$ can become less than unity at relatively
high temperatures and the difference between the strong and weak fields
is eliminated. Physically this crossover occurs because $\tau_\phi$
becomes less than $\tau_{ac}$, and the phase of the electron has been
randomized before it has time to absorb additional photons. In this case
the temperature dependence of $\overline{V_{AF}}$ is reduced to that expected 
for ``low'' microwave powers.

In the low-power limit, the temperature dependence of $\overline{V_{AF}}$
also depends on the relative value of $\tau_\phi$ and
$\omega^{-1}$.  In the high-frequency limit
($\omega^{-1} < \tau_\phi < \tau_f$) the temperature dependence 
of $\overline{V_{AF}}$ (in 1D) is obtained \cite{falko89a} from
Eq.~\ref{eq:Vdc4} (again substituting $L \rightarrow L_\phi$ and 
$\tau_f \rightarrow \tau_\phi$) 
\begin{equation}
\overline{V_{AF}}(T) 
\propto \tau_\phi^{-1/2} L_\phi^{2-d/2} 
\propto T^{-p/4}, ~~\omega \tau_\phi \gg 1.
\label{eq:tempdep4} \end{equation}
As the temperature is increased further the condition 
$\tau_\phi < \omega^{-1} < \tau_f $ holds and the temperature 
dependence of $\overline{V_{AF}}$ in the ``low-frequency'' limit 
(see Eq.~\ref{eq:Vdc4}) is given (in 1D) by 
\begin{equation}
\overline{V_{AF}}(T) \propto \tau_\phi L_\phi^{2-d/2}  
\propto T^{-7p/4}, ~~\omega \tau_\phi \ll 1.
\label{eq:tempdep1}
\end{equation}
In this low-frequency limit $V_{dc}$ is
expected to be a strong function of temperature in contrast to the
high-frequency limit. Therefore, studies of $V_{dc}$ over a wide
temperature range may not scale  according to a single power law
and may reveal evidence that the time scale of the microwaves
($\omega^{-1}$) has a strong influence on the behavior of a
mesoscopic system.

Analogous to UCF \cite{lee90,wash85}, thermal averaging 
of $n_T=k_BT/E_c$ uncorrelated energy intervals leads to a $T^{-1/2}$ 
reduction in magnitude of $V_{dc}$ at low powers. However, 
for $\hbar \omega > E_c$, the photon energy supersedes the correlation 
energy and the number of uncorrelated electron intervals is given by 
$ k_B T / \hbar \omega$.  For even larger photon energies 
(i.e. $\hbar \omega >  k_BT$) thermal smearing is circumvented altogether 
and $n_T=1$ \cite{falkonote}. 
Thermal averaging can introduce a second important length scale 
$L_T=\sqrt{hD/k_BT}$, or the thermal length, defined as the distance 
electrons differing in energy by $k_BT$ diffuse before their wave-functions 
are significantly out of phase. At high temperatures it is possible for
$L_T$ become smaller than $L_\phi$, where the fundamental mesoscopic 
length scale is determined by the smaller of the two.   

\subsection{The Photovoltaic Effect for Ring Samples}

The PV theory by Fal'ko and Khmel'nitskii \cite{falko89a}, as discussed
above, estimates the magnitude of the
aperiodic PV  fluctuations, $\overline{V_{AF}}$, for the case of a 
microjunction geometry. The theory has not yet been extended to include 
the $h/e$ PV oscillations, $V_{h/e}$, observed in samples with ring 
geometries \cite{bykov93,bartthesis}. In the low-frequency low-field 
limit we would expect the magnitude of
the aperiodic and $h/e$ PV oscillations to have similar power
dependences.  However, it is well known from the UCF studies
\cite{lee90} that the $h/e$ conductance oscillations are suppressed
more strongly by inelastic scattering than are the aperiodic
fluctuations. The $h/e$ PV oscillations are more
sensitive to inelastic scattering since the electron must traverse 
the entire circumference of the ring without having its phase randomized. In
contrast, the aperiodic fluctuations can result from relatively small
paths confined to the linewidth of the ring. 
As a result, the $h/e$ oscillations are suppressed 
exponentially according to a phase coherence survival probability, 
$\exp(-\pi r/L_\phi)$, for $L_\phi < \pi r$, where $r$ is the radius of 
the ring \cite{milliken87}.
In the case of the PV effect, we expect the dephasing due to the
microwave field to cause $\overline{V_{h/e}}$ and $\overline{V_{AF}}$
to scale quite differently as a function of microwave power. In the
absence of a quantitative theory, we predict that the 
power and temperature dependence of the $h/e$ PV oscillations is 
approximately given by
\begin{equation}
\overline{V_{h/e}} \sim E_{ac}^2\exp(-\pi r/ L_{eff}),
\label{eq:powexp}
\end{equation}
where $L_{eff} \approx \left[1/L_\phi^2 + 1/L_{ac}^2\right]^{-1/2}$ is
the effective phase coherence length.  In Eq.~\ref{eq:powexp} we see
that $\overline{V_{h/e}}$ scales quadratically with the microwave field
at low powers (where $L_{eff} \approx L_\phi$). As $L_{ac}$
becomes less than $L_\phi$, due to excessive electron-photon
scattering, the exponential suppression will dominate the square-law 
dependence. Therefore, a detailed experimental study of the power 
dependence of the PV effect in submicron-size rings should provide clear 
evidence of the dephasing effects of a microwave field in a mesoscopic 
system.

\section{EXPERIMENTAL SETUP}

The samples used in this study consisted of four submicron-diameter
Au rings fabricated using a scanning electron
microscope converted for electron beam lithography
\cite{bernstein93}. Although similar results were found for all four
rings, the data to be shown below pertain to only two of the
samples. Ring $\#1$ had an inner diameter of
$d \approx 3300 ~ {\rm \AA}$ and linewidth
$w \approx 550 ~{\rm \AA}$, and for ring $\#2$
$d\approx 4700 ~{\rm \AA}$ and $w \approx 700
~{\rm \AA}$.  Both rings were 200 ${\rm \AA}$ thick.
The sample dimensions were determined
to within $10 \%$ using the scanning electron microscope. Narrow
voltage probes, with the same linewidth as the ring, extended outwards
0.5 $\mu$m on opposite sides of the ring before widening to 8
$\mu$m. Therefore the wide voltage contacts were separated in length,
$L$, by  $ 1~\mu$m. We note that such wide voltage contacts may
act as antennas tending to strongly couple the sample and
the microwave radiation. The samples are effectively one-dimensional
since $L \gg w$. Based on the total sample resistance of 30 $\Omega$,
and the number of squares ($\sim$15), the sheet resistance was
estimated as $\approx 2.0 ~\Omega $ at $T=$ 4.2 K. The diffusion
constant was estimated from resistivity measurements to be
$D \approx 100 ~{\rm cm}^2$/s.

For all the PV data to be shown in this paper the microwave frequency
was 8.4 GHz corresponding to a microwave energy of
$\hbar \omega \approx 35 ~\mu$eV. This is smaller than the thermal
spread in energies $k_B T = 120 ~\mu$eV for our lowest measurement
temperature, 1.4 K, where $k_BT/\hbar \omega \approx 3.5$. For typical
disordered metallic samples \cite{lee90}, like the ones used in this
study, $L_\phi \sim 1 ~ \mu$m at 1 K and $D \sim 100 {\rm ~cm^2/s}$.
Accordingly, $E_c$ is usually of order 10 $\mu$eV yielding a
correlation temperature of $E_c/k_B=$0.1 K. For the temperatures used 
in this work, it is likely that the condition $E_c<\hbar \omega <k_BT$ 
holds and $V_{dc}$ should be affected by thermal smearing. Given these 
numbers, $\omega \tau_\phi \sim $ 5 for 10 GHz radiation, 
and Eq.~\ref{eq:Vdc4} should pertain to the high-frequency regime
for temperatures around 1 K.

The samples were mounted in a microwave cavity which had a resonant
frequency of $8.4 ~ \rm GHz$ for its TE 210 mode \cite{liuthesis}.
They were located at a  maximum of the electric field, with this
field directed in the plane of  the film.  The cavity was inside a
vacuum can which was usually filled  with liquid He to minimize Joule
heating of the sample by the microwave  field. This was all positioned
inside a superconducting solenoid  which provided a magnetic field
perpendicular to the  plane of the film.  The microwave field was
modulated at $150 ~ \rm Hz$, and the sample voltage was measured
with a lock-in amplifier using a  transformer coupled preamplifier.
This scheme allowed us to conveniently  measure the very small
signals which were encountered in the mesoscopic  PV effect. We
will refer to the voltage measured in this way as
$V_{dc}$ even though it was not a strictly dc measurement.

In order to compare the measured value of
$V_{dc}$ with the theory, the magnitude of the
microwave field was calibrated using the  procedure discussed
previously \cite{liuthesis}. The maximum output power of the
microwave generator was 1 mW, however, the maximum input power to the
microwave cavity was of order 0.25 mW due to attenuation in the
coaxial cable leading down the cryostat. By changing the frequency of
the microwaves (for fixed  input power) a resonant peak in $V_{dc}$
was observed which could then be used to estimate the
$Q$ of the cavity. The microwave field $E_{ac}$ was then calculated
using an estimated $Q$ value of 25 for the maximum microwave power.

The absolute magnitude of the microwave field could be in error by as
much as a factor of five due to possible inaccuracies in estimations
of the attenuation and the $Q$ value. However, the $\em
{relative}$ magnitude of the power (field) was accurately
determined from the attenuation setting on the
microwave power supply. The maximum attenuation of the input power
(0.25 mW) from the microwave generator, for which $V_{dc}$ was 
just indistinguishable from the instrumental noise, was -57 dB. 
The lowest power setting for a discernible PV signal was -52 dB. 
For the measured power dependence of 
$\overline{V_{dc}}$ to be discussed below, the power (field) settings 
ranged from -52 dB ($E_{ac}\approx 0.12~$V/m) to -10 dB 
($E_{ac}\approx ~$16 V/m).

\section{Measurements of the Photovoltaic Effect in Au Rings}

\subsection{Typical Results and Data Presentation}

\begin{figure}
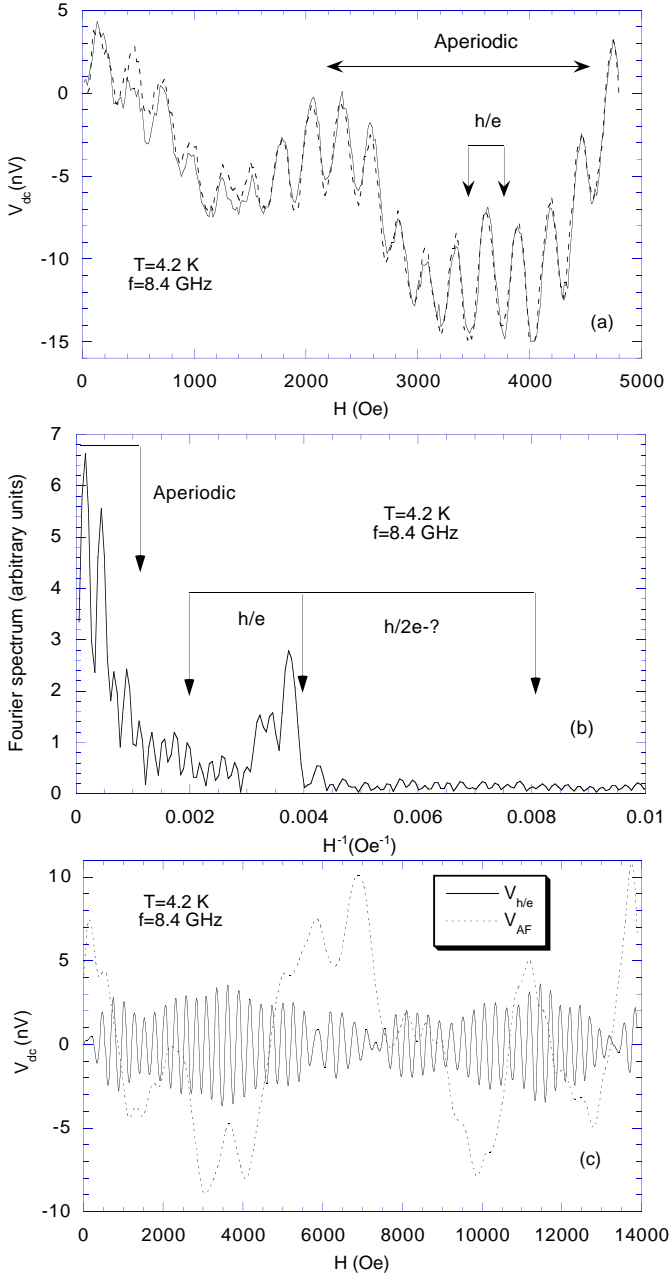

\caption{Aharonov-Bohm photovoltaic oscillations in Au ring $\#1$
($d=3300 ~\rm{\AA}$ and $w= 550~\rm{\AA}$) at $T=4.2 ~$K. (a) The solid
(dotted) line represents a field sweep with $H$ increasing (decreasing)
in magnitude. (b) The Fourier transform of $V_{dc}$ with arrows
indicating the various frequency ranges expected for the
aperiodic fluctuations ($AF$), $h/e$, and $h/2e$ oscillations based
on the measured ring geometry. (c) Using the digital filtering procedure
described in the text the aperiodic (dotted line) and the $h/e$
oscillations (solid line) may be clearly separated.}
\label{fig:ringdata}
\end{figure}

In Fig.~\ref{fig:ringdata}(a) we show typical
results for the magnetic field dependence of the PV effect for
ring $\#1$ at $T=$ 4.2 K. $V_{dc}$ was quite reproducible and
constituted a true `magnetofingerprint' where the solid (dotted)
line shows data for which the magnetic field was increasing
(decreasing) in magnitude.  As expected, the
data clearly reveal both the $h/e$ PV oscillation ($V_{h/e}$) and
the aperiodic  fluctuations ($V_{AF}$).
Based on the known geometry of the ring, the period of the $h/e$
oscillations was expected to be
$\sim$ 350 Oe. This is in agreement with the observed period of
approximately 300 Oe which can be obtained by inspection of
Fig.~\ref{fig:ringdata} (a) where we see roughly 3.5 oscillations over
the field range 3000 to 4000 Oe. The aperiodic correlation field,
$H_{c} (\sim \Phi_o/ L_\phi w) \approx 2500 $ Gauss, can also
be determined by inspection of Fig.~\ref{fig:ringdata}(a), where we
see two aperiodic fluctuations spaced over an approximate field range
of 5000 Oe. From the measured linewidth of the ring
$w=550 ~{\rm \AA}$, the phase coherence length can be estimated as
$L_\phi \approx 0.3~ \mu$m, which is roughly consistent with
previous results for similar samples and temperatures
\cite{milliken87}. The aperiodic and $h/e$ PV oscillations are
also easily observed in a Fourier
transform (FT) of the raw data as can be seen in
Fig.~\ref{fig:ringdata}(b). The $h/e$ frequency range
was 0.0025-0.0045 Oe$^{-1}$ which  corresponds to a field range
of 220-400 Oe. The $h/e$ peak in the FT is centered about
0.0035 Oe$^{-1}$ (or 285 Oe) which agrees with the value obtained
by inspection above.

We note however that the  $h/2e$ oscillations were conspicuously
absent from the  data and were not discernible from the Fourier
transforms.  This was surprising given the robust size of the
$h/e$ oscillations.  However, in analogy with the
UCF, the $h/2e$ oscillations are expected to be exponentially
suppressed in comparison to the $h/e$ oscillations.
Furthermore, the dephasing  effects of the microwave field would
further decrease the amplitude of the $h/2e$ oscillations, relative to
the $h/e$ oscillations, a fact which may explain their absence.

In order to compare with the theory
(Eqs.~\ref{eq:Vdc1}-\ref{eq:region3pow}) the rms
value of the PV fluctuations were computed.
The $h/e$ PV oscillations and aperiodic PV
fluctuations were separated by computing
the inverse FT after selecting a region surrounding one of
the peaks in the FT. The data after filtering is shown in
Fig.~\ref{fig:ringdata}(c) where we see that the frequency components
are clearly separated. The rms values of the
$h/e$ oscillations ($\overline{V_{h/e}}$) and aperiodic fluctuations
($\overline{V_{AF}}$) were then easily computed over a fixed magnetic
field range to compare with the theory.

\subsection{Power Dependence of the Photovoltaic Effect }

Before making a quantitative comparison of the power (field) dependence
of $\overline{V_{dc}}$ with the theory it is instructive
to first consider several qualitative features of the data. In
Fig.~\ref{fig:powerdep}(a) we show $V_{dc}$ for ring $\#2$ at 1.4 K
for microwave power levels differing  by roughly a factor of 1000.
For the moderate microwave field ($E_{ac} \approx 0.6~$V/m),
$V_{dc}$ exhibited $h/e$ oscillations superimposed upon the
slowly varying aperiodic fluctuations.  For the larger microwave 
field ($E_{ac} \approx 16~$V/m) it is seen that the
magnitude of the aperiodic  fluctuations clearly increased, however,
the $h/e$  oscillations were quenched and not distinguishable
above the noise.   At somewhat lower microwave levels
($E_{ac} \approx$ 0.12 V/m), as in Fig.~\ref{fig:powerdep}(b), the
aperiodic PV fluctuations were roughly the same size as the
$h/e$ oscillations. (Note the change in the voltage scale between
Figs.~\ref{fig:powerdep}(a) and (b).) For still lower microwave
levels the PV effect was difficult to distinguish above the noise.

Several semi-quantitative features of this data are clearly evident.
First, it is obvious that we are not in any type of linear response
regime. An increase in microwave power by a factor of 10,000 only
resulted in an increase in
$V_{AF}$ by roughly a factor of 35. This can be seen by comparing
Fig.~\ref{fig:powerdep}(b), where $V_{AF}(0.12 $ V/m) $\sim 1 $ nV, and
Fig.~\ref{fig:powerdep}(a) where $V_{AF}(16 $ V/m) $\sim 35$ nV.
This weak power dependence was to be expected from the logarithmic
power dependence predicted by the theory \cite{falko89a} 
(see Eq.~\ref{eq:region3pow}). The power dependence of the $h/e$ 
oscillations clearly shows 
that large microwave powers are causing dephasing since $V_{h/e} \approx 0$. 
The dephasing effects of the microwaves were also
evident through changes in the aperiodic correlation field  ($H_c \sim
\Phi_o/L_{eff} w$, where $\Phi_o$ is the flux quantum). In
Fig.~\ref{fig:powerdep}(a) this is clearly demonstrated where
$H_c$ is seen to increase by nearly a factor of seven as
$E_{ac}$ is changed by a factor of 30. The increase in the correlation
field at high microwave powers was not discussed
explicitly by the theory \cite{falko89a}, although it is likely due to a
decrease in $L_{eff}$ resulting from the dephasing nature
of the microwaves. 

The results of a quantitative study of the power dependence
of the PV effect at 1.4 K are shown in Fig.~\ref{fig:powerdep}(c).
Each data point represents the rms value of
$V_{dc}$ computed over a magnetic  field sweep from 0 to 8000 Oe.
Fig.~\ref{fig:powerdep}(c) contains data shown previously in
Figs.~\ref{fig:powerdep}(a) and (b), and for four other microwave
field values.  One aspect of these measurements that will limit a precise
comparison with the theory centers on the relatively  small number of
aperiodic fluctuations observed over the available magnetic field
range. Furthermore, more data could have been taken in smaller
increments of the microwave field. The basic power dependence of the
aperiodic PV fluctuations is clear however, where from
Fig.~\ref{fig:powerdep}(c) it can be seen that
$\overline{V_{AF}}$ (solid squares) grows
with the microwave field before saturating at high fields. In contrast
$\overline{V_{h/e}}$ grows more slowly with the microwave field before
becoming completely quenched for fields above about 8 V/m.

In Fig.~\ref{fig:powerdep}(c) we show the results of a fit (solid line)
to the logarithmic power dependence predicted for the aperiodic
fluctuations,
$\overline{V_{AF}}\sim b\ln (aE_{ac})$, as estimated by the theory (see
Eq.~\ref{eq:region3pow}). The parameters $a=6.6$ and $b=2.1$ were
determined from the fit. The data is in
agreement with a logarithmic power dependence over a wide range of
microwave field  strengths. Although we do not show the 
results of the fits here, both $\overline{V_{AF}}$ and
$\overline{V_{h/e}}$ were consistent with a quadradic field dependence 
(as expected in the linear response regime) for
$E_{ac}<$1 V/m. From  Fig.~\ref{fig:powerdep}(c) we see that the
transition between the linear response and the high-power regime occurs
for microwave fields above 1 V/m.  The theory predicts that this
transition should take place when the number of photons absorbed
(emitted) by the electron in time $\tau_f$ is greater than one [i.e.,
$\xi (=eE_{ac}L/\hbar \omega) > 1$]. Based on the microwave energy (35
$\mu$eV) and sample length ($L \sim$ 1 $\mu$m) the theory predicts a 
crossover for $E_{ac}\sim $35 V/m. Given this,  a fit
parameter of $a(= eL/\hbar \omega) \approx 1/35$, not 6.6,
should should have been obtained from the data. This discrepancy is hard
to reconcile with the data unless we have underestimated the magnitude
of the microwave field by at least an order of magnitude, or the power
absorbed by the sample by a factor of 1000.
From the observed power dependence of the
aperiodic fluctuations it is apparent that we are at least in agreement
the logarithmic power dependence predicted by the theory.

It is clear from Fig.~\ref{fig:powerdep}(c) that the power dependence
of the $h/e$ oscillations (triangles) is quite different from the
aperiodic fluctuations for microwave fields greater than 1 V/m. For
microwave fields greater than roughly 8 V/m, the $h/e$ oscillations
are completely quenched in contrast to the aperiodic fluctuations.
This behavior was expected from Eq.~\ref{eq:powexp} where it was
anticipated that the dephasing nature of the microwaves might give rise
to an exponential suppression at high powers. In
Fig.~\ref{fig:powerdep}(c)
the solid line results from a fit to following expression
$\overline{V_{h/e}} \approx bE_{ac}^2\exp(\pi r/L_{eff})$, where
$L_{eff} \approx \left[1/L_\phi^2 + 1/L_{ac}^2\right]^{-1/2}$ and
$L_{ac}^{-1} \approx aE_{ac}$, as expected from Eq.~\ref{eq:powexp}.
It was also assumed that $\pi r \approx L_\phi$ which is a reasonable
approximation at 1.4 K. We see that there is good agreement with
Eq.~\ref{eq:powexp}, although measurements at more values
of $E_{ac}$ are needed for a more stringent comparison. 

\subsection{Temperature Dependence of the PV Effect}

\begin{figure}
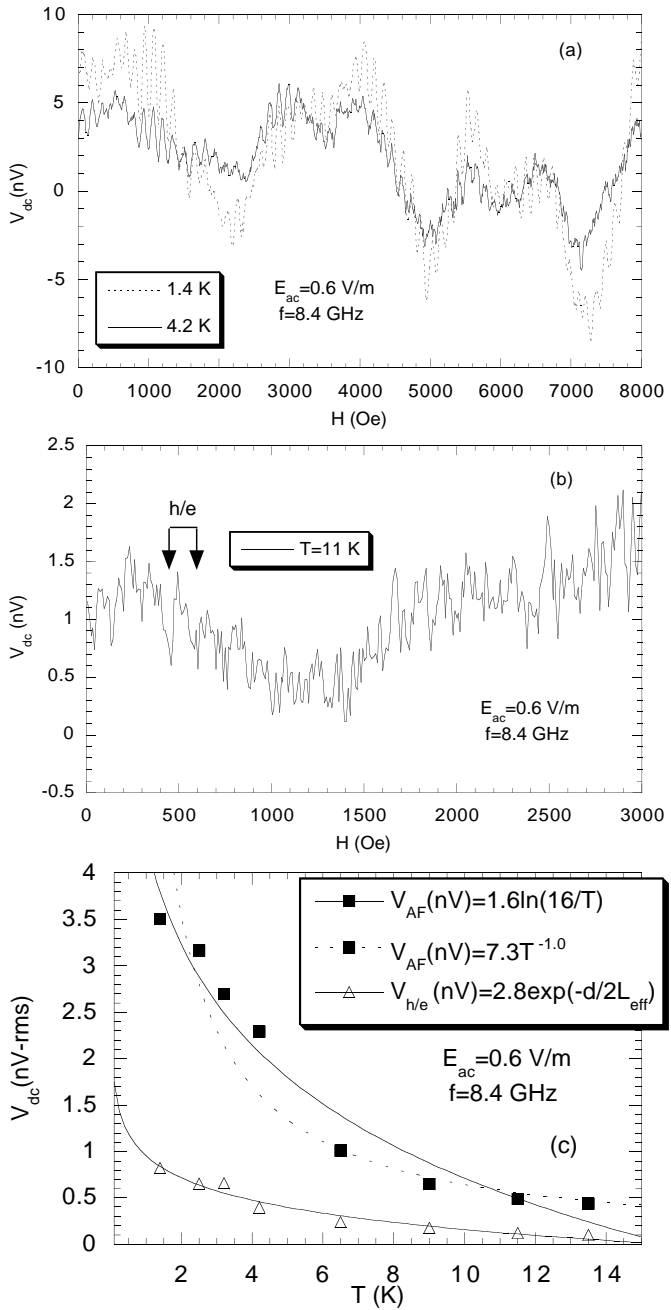

\caption{(a) PV results for Au ring $\#2$ ($d=4700 ~{\rm \AA}$) at two
different temperatures, 1.4 K (dotted line) and 4.2 K (solid line),
for the moderate microwave power ($0.6~$V/m).
(b) Close-up of data at 11 K.
(c) Plot of the rms values of the data shown in (a) and (b) and for 5
other temperatures.}
\label{fig:temp38}
\end{figure}

Fig.~\ref{fig:temp38}(a) shows raw PV data at two
temperatures for ring $\#2$ ($d=4700 ~{\rm \AA}$ and $w=700~{\rm\AA}$),
for the ``moderate'' microwave field ($E_{ac} \approx 0.6~$V/m) as shown in 
Fig.~\ref{fig:powerdep}(a). As expected, the magnitude of both the 
aperiodic and $h/e$ PV oscillations decreased as the temperature was 
increased. In Fig.~\ref{fig:temp38}(b) we take a closer look at data obtained
at $T=11~$K. The Aharonov-Bohm oscillations were visible but
were much smaller in comparison to the results
obtained at $T=1.4~$K. That the $h/e$ PV oscillations  persisted to such 
high temperatures was
surprising, since the analogous $h/e$ conductance oscillations are
usually difficult to detect at these temperatures ($>$ 5 K) for similar
samples \cite{wash85,milliken87}.

In Fig.~\ref{fig:temp38}(c) we have plotted the rms values of
both the aperiodic $\overline{V_{AF}}$ and
$h/e$ oscillations $\overline{V_{h/e}}$ for the raw data shown
in (a) and (b), and five other temperatures ranging from
$T=1.4~{\rm to} ~14~$K. The rms values were computed in the same
manner as discussed in connection with Fig.~\ref{fig:powerdep}(c).
From the theory \cite{falko89a}
(see Eqs.~\ref{eq:tempdep2}-\ref{eq:tempdep1}) we  would expect
$V_{AF}$ to decrease logarithmically as a function of temperature
in the high-power regime and according to a power-law
($\overline{V_{AF}} \sim T^{-p_1}$) in the low-power regime.
In the low-frequency limit, $p_1=7/4p+1/2$, while in the low-power
high-frequency regime the temperature dependence is much weaker
($p_1=p/4+1/2$), where the factor of 1/2 is due to thermal smearing. 
In Fig.~\ref{fig:temp38}(c)  we show the result of a
fit (solid line) to a logarithmic temperature dependence of the form
$\overline{V_{AF}} \sim a\ln(b/T)$ (see Eq.~\ref{eq:tempdep2}).
The fit yields reasonable  agreement for low temperatures ($<$ 4.2 K) 
although there is
some deviation from the logarithmic dependence at higher temperatures.
Given the scatter in the data, it is not entirely clear whether this
deviation warrants a  quantitative explanation. However, assuming the
deviation is real, the fact that the magnitude of the microwave field,
0.6 V/m, is at the transition between the low and high-power regimes
offers a possible explanation. In the high-power regime there are
multiple electron-photon interactions (i.e., $\xi > 1$) and $L_{ac} <
L_\phi$. As the temperature increases, $L_\phi$ becomes shorter
than $L_{ac}$ and temperature dependent inelastic processes
dominate. In this case, the logarithmic dependence gives way to the
power-law dependence. 

The dotted line in Fig.~\ref{fig:temp38}(c)
represents a fit to a power-law of the form $\overline{V_{AF}} \sim
AT^{-p_1}$, where the parameters $A=$ 7.4 and
$p_1=1.0$ were determined from the fit. The fit is seen to be in
good agreement with the data for temperatures greater than 6 K.
For lower temperatures ($< 4$ K), the agreement
is not satisfactory, even considering the scatter in the data.
This relatively weak temperature dependence is consistent with that
expected in the high-frequency limit, with $p=2$ due to
electron-phonon interactions. That we did not
observe the stronger temperature dependence expected for the
low-frequency regime (see Eq.~\ref{eq:tempdep1}) indicates that the
frequency of the microwaves replaced $\tau_\phi$ as the dominant
time scale in a mesoscopic system.
The weak temperature dependence exhibited by the aperiodic
PV fluctuations is consistent with previous results \cite{liu92},
although those authors concluded that the observed temperature dependence was
weaker than expected from the theory \cite{falko89a}. Our analysis resolves
this apparent discrepancy. Unfortunately it was not possible to
determine the temperature dependence for microwave fields smaller than
0.6 V/m, since for this value of the microwave field $V_{dc}$ was quite
small for temperatures greater than 1.4 K.

We also studied the temperature dependence of the corresponding
$h/e$ PV oscillations, $\overline{V_{h/e}}$, shown as the
triangles in Fig.~\ref{fig:temp38}(c). From Eq.~\ref{eq:powexp}
we would expect a weak exponential temperature dependence since
the microwaves contribute to an effective  phase coherence
length, $L_{eff} \approx \left[1/L_\phi^2+1/L_{ac}^2\right]^{-1/2}$,
acting to weaken the temperature dependent length scale $L_\phi$. 
In Fig.~\ref{fig:temp38}(c) the result of a fit to the $h/e$
data of the form $\exp(-\pi r/ L_{eff}) \sim \exp[-(T^p+b)^{1/2}]$
is shown as the solid line, where it was assumed that $\pi r \approx
L_\phi$ at 1.4 K and $b=L_\phi^2/L_{ac}^2$. The fit is quite reasonable 
over the entire range of temperature.  We note that the data is not at 
all consistent with an exponential suppression of the form 
$\exp(-\pi r/L_\phi) \sim \exp(-T)$ and do not
show the results of such a fit here. For this reason, it seems
likely that the microwave field is the dominant source of inelastic
scattering as opposed to electron-electron or electron-phonon 
interactions. We note that this weak exponential suppression closely 
mimics the logarithmic [$\ln(1/T)$] and power-law ($T^{-1}$)
temperature dependences observed for the aperiodic fluctuations. 
A deeper explanation for the observed temperature dependences awaits 
more extensive measurements and a full theory of the PV effect to include 
the case of a ring geometry.

The temperature dependences for even larger microwave fields ($E_{ac} >$
0.6 V/m) were also studied. For $E_{ac}\approx$5.0 V/m the
temperature dependence of $V_{AF}$ and $V_{h/e}$ become progressively
weaker, although similar to Fig.~\ref{fig:temp38}(c), and we do not 
show the data here.  
For the largest microwave field (16 V/m) used in this study 
(see Fig.~\ref{fig:powerdep}), the temperature dependence was 
quite weak. As can be seen, $\overline{V_{AF}}$ showed very
little change as the temperature was increased from 1.4 to 16 K. 
For this relatively high microwave field the Aharonov-Bohm oscillations 
are completely quenched since we are in the high-power regime. 
In Fig.~\ref{fig:highpow}(b)
we show the results  of a quantitative comparison of the theory,
$\overline{V_{AF}} \sim a\ln(b/T)$ where the data does not quite agree 
with the logarithmic temperature dependence predicted by the theory. 
The data for $V_{AF}$ at
1.4 K in Fig.~\ref{fig:highpow}(a) was shown previously in
Fig.~\ref{fig:powerdep}(a) where we saw a large increase in
$H_c$ as a function of the microwave field. It is interesting to
note that despite an increase in temperature of 15 K there was very
little change in the correlation field $H_c$. It would therefore
appear that the microwaves, in the moderate to high-power limit, are
a much more effective source of dephasing than temperature dependent 
inelastic scattering mechanisms. 

\section{Discussion}

We have presented a systematic study of the PV effect in
submicron-diameter Au rings as a function of microwave power and
temperature. We have clearly observed the quenching of the 
$h/e$ PV oscillations due to the dephasing effect of a microwave
field. Such dephasing may also explain the increase in $H_c$ for the
aperiodic fluctuations, although alternative
explanations for the observed behavior should be discussed.     
Our results for the power dependence of the PV effect are
are analogous to previous studies of the I-V characteristics
of UCF \cite{larkin86} where the magnitude of voltage
fluctuations were measured as a function of the dc-voltage bias,
$V_{b}$ \cite{haucke90,webb88}. These studies showed
that both the $h/e$ and the aperiodic conductance oscillations
suffered from voltage averaging and scaled as a weak power-law, 
$V_{b}^{1/2}$, for $V_{b} > V_c$, where 
$V_c = E_c/e$ is the correlation voltage. This was found for dc-fields of 
order $E_{dc}(=V_{b}/L) \sim$1 V/m, which is the same order of magnitude 
as the microwave fields used in this work. 
One analogous feature of the UCF results \cite{haucke90}
was that $H_c$ increased as a function of the dc-voltage bias across
the sample for $V_b > V_c$ due to the introduction of a dc-voltage
averaging length, $L_E=[\hbar D/ eV]^{1/2}$ (or $[\hbar D/
eE_{dc}]^{1/3}$).

\begin{figure}
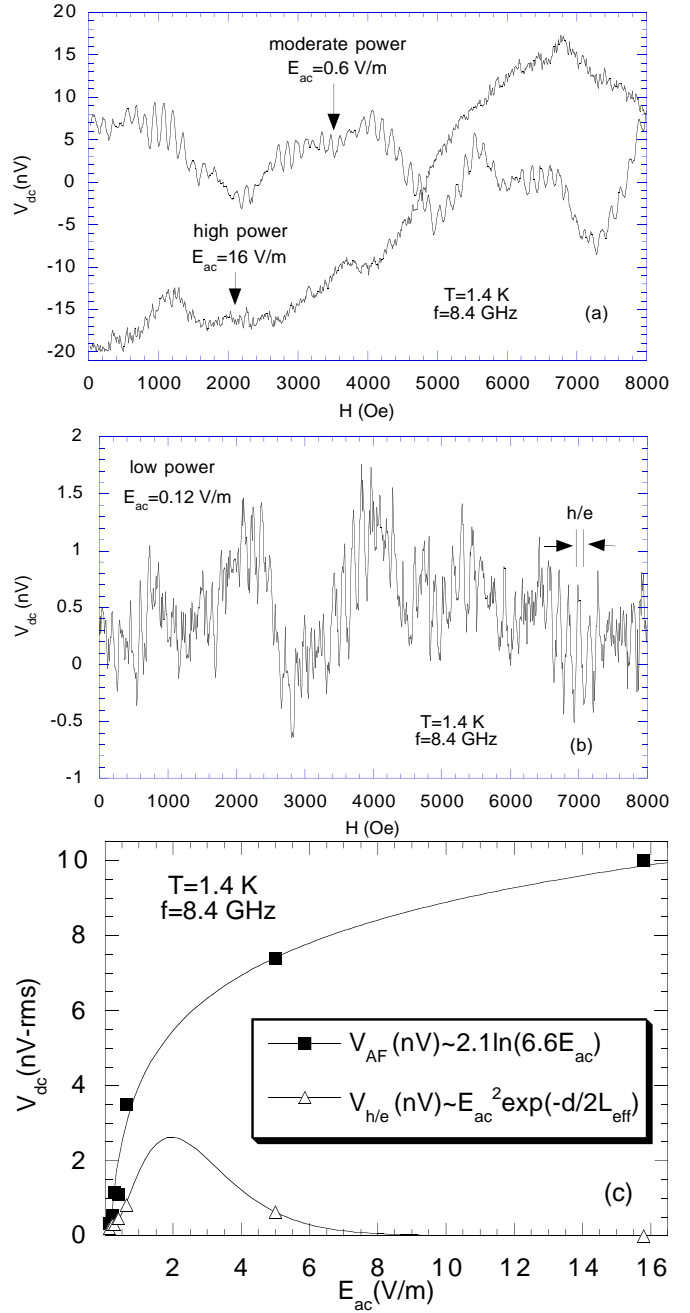

\caption{(a) $V_{dc}$ as a function of $H$ at $T=$1.4 K for a
moderate ($E_{ac} \approx 0.6~$V/m) and a high ($E_{ac} \approx 16~$V/m)
microwave power for the $d=4700 ~\rm{\AA}$ Au ring.
Two features of the power
dependence are clearly visible: The correlation field $H_c$ increases as
a function of the microwave power and the $h/e$ oscillations are greatly
suppressed at high powers. (b) For the lowest microwave field
($E_{ac} \approx 0.12~$V/m) the $h/e$ oscillations are almost the same
size as the aperiodic fluctuations. Note the change in the voltage scale
between (a) and (b). (c) $\overline{V_{AF}}$ (squares) and
$\overline{V_{h/e}}$ (triangles) as a function of the microwave field
$E_{ac}$.}
\label{fig:powerdep}
\end{figure}

In the case of the PV effect it is possible for both $L_E$ and $L_{ac}$
to enter the problem depending on the microwave field strength and
frequency. $L_E$ was predicted by the PV theory
\cite{falko89a} to become significant when
$\omega^{-1}$ is much larger than $\tau_f$ and pertains to the
low-frequency limit. In this case the magnitude of microwave
field can be considered to be essentially constant in time and many
features of the PV effect would be quite analogous to UCF. 
It is difficult to tell
whether the observed increase in $H_c$ in our PV studies (as shown in
Fig.~\ref{fig:powerdep}) resulted from an increase in $L_E$
(i.e., in the dc or low-frequency limit) or
$L_{ac}=\hbar \omega / eE_{ac}$ (strong high-frequency field) since
both are inverse functions of the electric field. 
However, the fact that the $h/e$ PV oscillations are clearly quenched 
for microwave fields larger than  5 V/m, 
as shown in Fig.~\ref{fig:powerdep}(c), is a strong indication that
the microwaves are dephasing. 
In contrast, the $h/e$ conductance fluctuations were not quenched 
in response to a dc-voltage bias and grew instead according to a power-law.

For small samples at low temperatures it is also entirely possible for a
finite voltage bias to heat the electron temperature above the
lattice temperature. This increases the amount of inelastic
scattering thus reducing $L_\phi$. Heating effects due to the microwave
field are difficult to rule out since they also grow with microwave
field strength just as the dephasing length $L_{ac}$ 
(see Eq.~\ref{eq:tau_ac}).  Roughly speaking, heating effects are
expected to occur  as the energy gained from the voltage bias,
$eV_{b}$, is of order $k_BT$. A quantitative treatment of the flow of
energy from the sample to the substrate, and on to the He bath, is
certainly more involved \cite{kuchar91}, although this provides a rough
upper bound. For even our lowest measurement temperature (1.4 K),
where $k_BT = 120 ~\mu$eV, a dc electric field around 120 V/m in
our 1 $\mu$m long wire would be necessary to cause significant heating.
According to our calibrations, the microwave fields used in this study
were much smaller than this, although it is possible that we have
underestimated the magnitude of the microwave field or the amount of
power absorbed by the sample. We note that studies of the I-V
characteristics of conductance fluctuations in micron-size Au rings
\cite{haucke90,webb88} observed some mild heating effects for
dc-voltage biases $\sim $ 1 V/m. However, these measurements
were performed at much lower temperatures ($\sim $ 50 mK) in contrast
to our relatively high temperatures ($ \sim 4$ K).  We  concur that
it is difficult  to ever completely rule out the possibility of heating
effects. If the electron temperature was increased due to heating from
the microwaves, this would increase the electron-electron inelastic
scattering rate and decrease $L_\phi$, the precise effect we are
claiming is causing the increase in $H_c$ and the suppression of the
$h/e$ PV oscillations.

Assuming that the observed suppression of the $h/e$ PV oscillations in
Fig.~\ref{fig:powerdep} is  not due entirely to heating, it is
interesting to interpret the power dependence as resulting
from multiple electron-photon scattering events. The number of such
scattering events is given by the number of
photons absorbed by the electron as it diffuses to the voltage contacts,
as measured by the parameter $\xi$, see Eq.~\ref{eq:photonabs}. Given
this, the results of Fig.~\ref{fig:powerdep}(c) may be interpreted as
resulting  from the dephasing nature of electron-photon interactions.
The PV effect may allow one to quantify, in a controlled manner, the 
number of electron-photon scattering events necessary to completely 
randomize the phase of the
electron.   However, a more precise estimate of the magnitude of the
microwave field, and its coupling to the sample, would be needed to
make a direct comparison. It is hoped that future theoretical and
experimental efforts will be able to accurately address these
fundamental and important aspects of the PV effect.

We wish to thank K. Hong, J. Liu, and V. I. Fal'ko for helpful
discussions. This work was supported by the NSF through grants
DMR-9220455 and DMR-9531638. G. H. Bernstein acknowledges partial
support from the NSF.

\vspace{0.25in}

$^*$Present address: Department of Physics, University of Maryland,
College Park, MD 20742-4111. E-mail address: rbartolo@wam.umd.edu

$^\dagger$Present address: SGS-Thomson Microelectronics, Inc.,
Carrollton, TX 75006.

\begin{figure}
\caption{$V_{dc}$ as a function of $H$ for Au ring $\#$2 
($d= ~{\rm 4700 ~\AA}$ and $w=700 ~{\rm \AA}$) in the high field
limit ($E_{ac} = 16~$V/m) for two different temperatures.
(a) $V_{AF}$ at $1.4~$K (solid line) and 16 K (dotted line).
Note that for this relatively high microwave
power the Aharonov-Bohm oscillations are completely quenched.
(b) The rms values of the above data where the solid line is a fit
to the logarithmic temperature dependence,
$\overline{V_{AF}} \sim b\ln(a/T)$, as discussed in the text.}
\label{fig:highpow}
\end{figure}


\begin{references}
\bibitem{falkonote2}
We note that the theory \cite{falko89a} did not replace $L$ with
$L_\phi$, a point which appeared to be overlooked.
\bibitem{falkonote}
Note that the theory \cite{falko89a} incorporates the effects of
thermal  smearing by scaling the magnitude of $V_{dc}$ by the
parameter $t$ which is the inverse of $n$ as defined in the text.
It is also important to note that the PV theory \cite{falko89a} assumes
that $E_c$ is independent of temperature. In UCF theory the
correlation energy is temperature dependent
($E_c \sim \hbar/\tau_\phi$) and can be a strong function of temperature
when $\tau_\phi (T)< \tau_f$. For high temperatures the effects of
thermal broadening may not be significant since $E_c$ can be quite
large. This would eliminate the factor $T^{-1/2}$ in
the temperature dependence of $V_{dc}$ and act to weaken the temperature
dependence from that expected from the theory.
\end{references}
\end{document}